\documentclass[a4paper,11pt]{article}
\pdfoutput=1 % if your are submitting a pdflatex (i.e. if you have
\usepackage{jcappub} % for details on the use of the package, please
\usepackage[T1]{fontenc} % if needed

\renewcommand{\l}{\left(}
\renewcommand{\r}{\right)}

\title{Freeze-out of baryon number in low-scale leptogenesis}

\author{S. Eijima,}
\author{M. Shaposhnikov,}
\author{I. Timiryasov}
\affiliation{Institute of Physics, Laboratory for Particle Physics and Cosmology,\\
\'{E}cole Polytechnique F\'{e}d\'{e}rale de Lausanne, CH-1015 Lausanne, 
Switzerland}

\emailAdd{Shintaro.Eijima@epfl.ch}
\emailAdd{Mikhail.Shaposhnikov@epfl.ch}
\emailAdd{Inar.Timiryasov@epfl.ch}

\abstract{
Low-scale leptogenesis provides an economic and testable description
of the origin of the baryon asymmetry of the Universe.
In this scenario, the baryon asymmetry of the Universe is reprocessed from the lepton asymmetry
by electroweak sphaleron processes. 
Provided that sphalerons are fast enough to maintain equilibrium,
the values of the baryon and lepton asymmetries are related to each other.
Usually, this relation is used 
to find the value of the baryon asymmetry at the time of the sphaleron freeze-out.
To put in other words, the formula which is valid only
when the sphalerons are fast,
is applied at the moment when they are actually switched off.
In this paper, we examine the validity of such a treatment.
To this end, we solve the full system of kinetic equations
for low-scale leptogenesis. This system includes equations describing
the production of the lepton asymmetry in oscillations of right-handed neutrinos,
as well as a separate kinetic equation for the baryon asymmetry.
We show that for some values of the model parameters,
the corrections to the standard approach are sizeable.
We also present a feasible improvement to the ordinary procedure, which accounts for these
corrections.
}

\begin{document}
\maketitle
\flushbottom

\section{Introduction} % (fold)
\label{sec:inroduction}

The existence of the baryon asymmetry of the Universe (BAU)
is among  the most challenging problems of modern cosmology.
Leptogensis is an appealing solution to the BAU problem
arising as a by-product of the see-saw mechanism of neutrino mass generation.

In the original---high-scale leptogenesis---scenario \cite{Fukugita:1986hr},
the lepton asymmetry originates from $CP$ violating decays of heavy 
(well above the electroweak scale) Majorana fermions.
By low-scale leptogenesis we mean the scenario of baryogenesis via right-handed neutrino 
oscillations which was proposed in ref. \cite{Akhmedov:1998qx} 
and further elucidated in refs. \cite{Asaka:2005pn,Shaposhnikov:2008pf}.
In both the high-scale and low-scale leptogenesis scenarios, the lepton asymmetry 
is reprocessed into the baryon asymmetry
by electroweak sphaleron processes \cite{Kuzmin:1985mm}.
If fast enough, these processes 
keep the baryon asymmetry $B$ in equilibrium with the
lepton asymetries $L_\alpha$ ($\alpha = e, \mu, \tau$)
\begin{equation}
	B(T) = \chi(T)\sum_\alpha \Delta_\alpha(T),
	\label{standard_formula}
\end{equation}
where the combination $\Delta_\alpha = L_\alpha - B/3$ is not affected by sphalerons,
and $\chi(T)$ is a coefficient of order of $1$; for instance,
well inside the  symmetric phase it is equal to $-28/79$~\cite{Khlebnikov:1988sr}.

The equilibrium\footnote{Unless otherwise stated, in what follows by equilibrium
  we mean equilibrium with respect to the sphaleron processes.}
 formula \eqref{standard_formula}
 for the baryon asymmetry \emph{is valid as long as the sphaleron rate exceeds 
 that of the lepton asymmetry production during all stages of BAU generation.}
 \newline
This is the case for high-scale leptogenesis, $T \gg 10^{8}$~GeV.
 However, this is definitely not the case for low-scale leptogenesis.
At temperatures around $T_{sph}\simeq131.7$~GeV (see ref. \cite{DOnofrio:2014rug} for the most recent computation of the freeze-out temperature), sphalerons become inoperative
while the lepton asymmetry is still being produced. One usually calculates
the eventual BAU by taking the value of $B$ at $T=T_{sph}$ from eq. \eqref{standard_formula}
even though the relation is not applicable at this temperature.

An  argument to justify the approach based on the use
of eq. \eqref{standard_formula}
in the low-scale leptogenesis
is that the sphaleron rate
drops exponentially with temperature \cite{Burnier:2005hp,DOnofrio:2014rug}.
This implies that sphalerons are not fast enough to maintain equilibrium
 only in a short transition region.
Nevertheless, the  rate of asymmetry generation might be large
in this transition region.
Therefore, it is not clear in advance how accurate the usual approach is.
This motivates us to study the dynamics of the $B$ freeze-out in detail.
To this end, we include a separate kinetic equation for 
$B(T)$ into the system of kinetic 
equations describing the production of lepton asymmetry in oscillations of 
right-handed neutrinos (heavy neutral leptons in terminology of the Particle Data Group, 
HNLs for short).

This work is organised as follows. In sections~\ref{sec:structure_of_kinetic_equations} 
and \ref{sec:susceptibilities},
we discuss how one has to modify the kinetic equations relevant for low-scale leptogenesis 
in order to account for 
various processes in the plasma. We describe two different approaches
to the computation of the BAU: the usual one, which assumes an instant freeze-out of 
the sphaleron processes, and 
a more general one with a separate equation for the baryon number. In section~\ref{nuMSM}, we  present the  kinetic equations for low-scale leptogenesis.
In section~\ref{sec:analysis_of_kinetic_equations}, we solve the equations in both 
approaches and analyse the results. 
In section~\ref{sec:improvement_of_the_instant_freeze_out_approach}, we propose
a feasible improvement of the approach with instant freeze-out of the BAU.
We conclude in section~\ref{sec:conclusions}. Our notations are fixed in 
appendix~\ref{sec:parametrization_of_yukawa_couplings}.

% section inroduction (end)

\section{Structure of kinetic equations} % (fold)
\label{sec:structure_of_kinetic_equations}

We are interested in models of low-scale leptogenesis. The kinetic equations
(see, e.g. \cite{Asaka:2005pn})
describe the generation of lepton asymmetries due to the interactions of active neutrinos
with HNLs.
The generated asymmetries will be redistributed among other Standard Model (SM) species 
due to various fast processes in the plasma.
In this section, we discuss a general procedure accounting for this redistribution.
We introduce a separate kinetic equation for a baryon number,
which allows us to trace the baryon asymmetry accurately.
In order to clarify the interplay between different physical processes,
we do not specify the form of the kinetic equations.
Recently, there was significant progress made by different 
groups \cite{Hernandez:2016kel,Drewes:2016gmt,Eijima:2017anv,Ghiglieri:2017gjz}.
The equations obtained in these works exhibit the same structure, 
but are different in details.
We leave a thorough analysis and comparison of these equations for future work.
In section \ref{nuMSM}
we will apply the results derived here to 
the kinetic equations obtained in ref. \cite{Eijima:2017anv}.

The physical system under consideration is very complicated. 
One can achieve a considerable simplification by separating fast and slow variables
and subsequently integrating out the fast variables (see, e.g., book \cite{zubarev1996statistical}
and references therein).
Right-handed neutrinos couple to the SM particles only through small Yukawa couplings.
Suppose for the moment, that these Yukawa interactions are switched off. In this case the charges
$\Delta_\alpha=L_\alpha-B/3$ are strictly conserved.
If the Yukawa couplings have small but non-zero values, the
number densities $n_{\Delta_\alpha}$ become slowly varying quantities.
The smallness of the Yukawa couplings implies that all variables
associated with HNLs are slow as well. We will generically denote  these variables,
namely, number densities of HNLs,
correlations between two HNLs, and analogous for anti-HNLs
(for Majorana fermions, one can associate different helicities with particles and anti-particles)
as $\mathbf{n}_{N}$.
At the temperatures relevant for the low-scale leptogenesis all
SM species are in thermal equilibrium.
This implies that chemical potentials of leptons  are equal within every generation:
${\mu_{\nu_\alpha} = \mu_{e_{L, \alpha}} = 
\mu_{e_{R, \alpha}} = \mu_\alpha }$.
The common chemical potentials to lepton numbers $\mu_\alpha$ vary
due to the Yukawa interactions with HNLs and due to the sphaleron processes.

The kinetic equations can be schematically written in the following way
\begin{equation}
	\begin{aligned}
	\dot{n}_{\Delta_\alpha} &= f_\alpha(\mathbf{n}_N, \mu_\alpha), \\
	\dot{\mathbf{n}}_{N} &= g_I(\mathbf{n}_N, \mu_\alpha).
	\end{aligned}
	\label{simple_sys_3}
\end{equation}
On the l.h.s. of the kinetic equations we use $n_{\Delta_\alpha}$ which can be changed
only due to interactions with HNLs. On the r.h.s. we keep using chemical potentials $\mu_\alpha$
to lepton numbers. 

The neutrality of the electroweak plasma implies a non-trivial relation between
the chemical potentials $\mu_\alpha$  and the asymmetries $n_{\Delta_\alpha}$
in the above equations. 
In the limit of small chemical potentials,
one can establish the following relation by means of equilibrium thermodynamics
\begin{equation}
	\mu_\alpha = \omega_{\alpha \beta}(T) n_{\Delta_\beta} + \omega_B(T) n_B,
	\label{susc_def}
\end{equation}
where $\omega$ are the 
susceptibility matrices,\footnote{To be more precise, here we deal  with the inverse susceptibility matrices,
see, e.g. ref. \cite{Ghiglieri:2015jua}. Note also that we have changed the definition of 
$\omega_{\alpha \beta}$ as
compared to ref. \cite{Eijima:2017anv},
by absorbing $T^2/6$.}
which we will calculate  in the next section.
Note that in eq. \eqref{susc_def}, we have included  the baryon number density as a separate variable.
This actually implies that sphaleron processes have to be treated as slow,
while the redistribution of charges in the plasma described by the 
functions $\omega_{\alpha \beta}$ and
$\omega_B$ is a fast process.\footnote{ 
Indeed, the rate of sphaleron processes is parametrically smaller than that of the other processes 
in the hot plasma.}
Therefore we have to introduce a separate kinetic equation for the baryon number density.
This was  derived in ref. \cite{Khlebnikov:1988sr} and
can be conveniently written in the following form
\cite{Burnier:2005hp}
\begin{equation}
	\dot{n}_B = -\Gamma_B (n_B-n_{B^{eq}}),
	\label{kine_ec_B}
\end{equation}
where for the three SM generations
\begin{equation}
   \Gamma_B= 3^2\cdot \frac{869 + 333 (\langle \Phi \rangle/T)^2}{792 +306 (\langle \Phi \rangle/T)^2}
    \cdot \frac{\Gamma_{diff}(T)}{T^3} , 
   \label{Gamma_B}
\end{equation}
and the equilibrium (with respect to sphalerons and other processes in neutral plasma) 
value of the baryon asymmetry is given by
\begin{equation}
	n_{B^{eq}}=- \chi(T)  \sum_\alpha n_{\Delta_\alpha} , \quad \chi(T) =
	 \frac{4 \left(27 (\langle \Phi \rangle/T)^2+77\right)}{333 (\langle \Phi \rangle/T)^2+869}.
	 \label{B_eq}
\end{equation}
Here  $\langle \Phi \rangle$ is the Higgs vacuum expectation value, 
which is equal to $246$~GeV at zero temperature. In a pure gauge theory,
the Chern-Simons diffusion rate  $\Gamma_{diff}(T)$ entering equation \eqref{Gamma_B},
was computed
in large-scale lattice simulations with the experimentally measured value
of the Higgs mass \cite{DOnofrio:2014rug}. It can be parametrized as (we do not include errors)
\begin{equation}
	\Gamma_{diff}\simeq
	\begin{cases}
	T^4\cdot\exp\left(-147.7+0.83T/\mbox{GeV}\right),&\mbox{in the Higgs phase},\\
	T^4 \cdot 18 \,\alpha_W^5, &\mbox{in the symmetric phase}.
	\end{cases}
	\label{G_diff}
\end{equation}
Note in passing, that the temperature of the sphaleron freeze-out
is defined as $\Gamma_B(T_{sph}) = H(T_{sph})$, where $H(T) = T^2/M_0$
and ${M_0(T) = \sqrt{90/g^*(T)}M_{\text{Planck}}/\pi}$.
In ref.~\cite{DOnofrio:2014rug}, it was found to be 
${T_{sph}=(131.7\pm2.4)}$~GeV. Throughout this paper we will use the central value ${T_{sph}=131.7}$~GeV.

As one can see from eqs. \eqref{Gamma_B} and \eqref{G_diff}, $\Gamma_B(T)$
drops sharply with  temperature in the Higgs phase. Therefore, it is reasonable to assume
that sphalerons maintain ${n_B(T) = n_{B^{eq}}(T)}$ at ${T>T_{sph}}$,
and after they are switched off instantly, so that
$n_B(T) = n_{B^{eq}}(T_{sph})$ at all temperatures below $T_{sph}$.
This, in fact standard, assumption allows one to simplify the computations,
since eq. \eqref{kine_ec_B} is no longer needed. For the sake of brevity,
in what follows we will refer to this approach with instant freeze-out, 
 as \textbf{Approach 1}. The alternative \textbf{Approach 2},
is a solution of the system \eqref{simple_sys_3} together with the
kinetic equation for the baryon number density \eqref{kine_ec_B}.
This is more general and will be used to examine
the assumption of  instant freeze-out.
Systems of the type \eqref{simple_sys_3}, 
 usually contain many different time scales and therefore
 exhibit a stiff behaviour. 
An obvious drawback of \textbf{Approach 2} is that we add a new
fast time scale $\Gamma_B^{-1}$ to an already stiff system. However,
this is the price  to pay  in order to be able to track $n_B(T)$.

The system of equations~\eqref{simple_sys_3} supplemented by relations \eqref{susc_def}
and eq. \eqref{kine_ec_B} is the main subject of this paper.
To the best of our knowledge the behaviour of this system has never been studied in the 
literature.\footnote{The behaviour of the BAU governed by eq.~\eqref{kine_ec_B} with a 
temperature-dependent
source of lepton asymmetry was studied in section 5 of ref.~\cite{DOnofrio:2012phz}.}

% section structure_of_kinetic_equations (end)

\section{Calculation of susceptibilities} % (fold)
\label{sec:susceptibilities}
As we have mentioned above,
the relations between the densities of the conserved charges and the corresponding 
chemical potentials are modified
by the requirement of neutrality of the plasma.
In this section,
we recall the derivation of susceptibility matrices introduced in eq. \eqref{susc_def}.

The redistribution of charges in the electroweak plasma can be described in different ways.
For example, one can relate different chemical potentials
by imposing equilibrium conditions that hold at a given temperature. 
This allows one to establish relations analogous to eq. \eqref{susc_def}, but
with temperature independent coefficients, see, e.g. \cite{Nardi:2006fx}.
Another method relies on the calculation of the thermodynamical potential density
(which is equal to minus the pressure) $\Omega(\mu, T)/\mathcal{V}$. 
As was shown in ref. \cite{Khlebnikov:1996vj},
this method allows to account for the temperature dependence of Higgs vacuum expectation
value.
This is important for  low-scale leptogenesis since it  allows one
to address  both the symmetric
and Higgs phases simultaneously.

The way to compute $\Omega(\mu, T)$ for systems with  Higgs mechanism 
and in the presence of conserved global charges
was introduced in ref. \cite{Khlebnikov:1996vj}. 
 Here we outline the recipe from ref. \cite{Khlebnikov:1996vj}.
It is important that the temporal components of the gauge fields play role 
of chemical potentials. Therefore, we denote them as
\begin{equation}
	\mu_Y\equiv i g_1 B_0, \quad \mu_T \equiv i g_2 A_0^3.
\end{equation}
For each particle species, we introduce the chemical potential
\begin{equation}
	\mu_I = \sum_i g^i_I \mu_i+ Y_I \, \mu_Y + T_{3, I} \,\mu_T,
\end{equation}
where $I$ enumerates the species and $i$ enumerates the conserved global charges 
$g^i_I$.
The one-loop potential density is constructed as
\begin{equation}
	\Omega(\mu, T)/\mathcal{V} = V_T(\langle \Phi \rangle) - V_0(\langle \Phi \rangle) - T^2 \sum_I \eta_I \mu_I^2,
\end{equation}
with  the temperature dependent Higgs potential $V_T$,  the overall volume $\mathcal{V}$, and
 $\eta_I = 1/12$ for massless fermions, while for massless bosons  $\eta_I = 1/6$.
Corrections coming from the masses of particles can be  accounted for 
(see, e.g., ref. \cite{Ghiglieri:2016xye}),
but they are not important for our considerations here.

As we have mentioned in the previous section, 
we assume that the sphaleron processes and interactions with
HNLs are slow as compared to the interactions in the electroweak plasma.
Therefore, we consider that the baryon  and individual lepton numbers are
conserved separately. We introduce the chemical potentials
$\mu_B$  and $\mu_\alpha$ corresponding to $B$ and $L_\alpha$, respectively.
Thus, the thermodynamical potential is a function of 
${\mu_B, \mu_\alpha, \mu_T, \mu_Y}$ and $T$
\footnote{Note that we have included the contributions of all the 
SM species (three generations of  left- and right-handed charged leptons and quarks,
 left-handed active neutrinos,
as well as  gauge bosons)
 to the  potential \eqref{Omega}. This is valid as long as we are interested in 
relatively high temperatures, which is the case for the present study.
}
\begin{equation}
	\begin{aligned}
	\Omega(\mu, T) / \mathcal{V} &= \frac{1}{24}  \left(8 T^2 \mu _B^2+8 T^2 \mu _B \mu _Y+6 \mu _1^2 T^2+6 \mu _2^2 T^2+6 \mu _3^2 
	T^2+22 T^2 \mu _T^2+22 T^2 \mu_Y^2-\right.\\
	& \left. 8 \mu _1 T^2 \mu _Y-8 \mu _2 T^2 \mu _Y-8 \mu _3 T^2 \mu _Y+3 \langle \Phi \rangle^2 \mu _T^2-6  \langle \Phi \rangle^2 
	\mu _T \mu _Y+ 3 \langle \Phi \rangle^2 \mu_Y^2\right).
\end{aligned}
	\label{Omega}
\end{equation}
The number densities of the conserved charges are given by
\begin{equation}
	-\frac{\partial (\Omega / \mathcal{V})}{\partial \mu_\alpha} = n_\alpha, \quad 
	-\frac{\partial (\Omega / \mathcal{V})}{\partial \mu_B} = n_B,
	\label{LandB}
\end{equation}
while the neutrality conditions read
\begin{equation}
	\frac{\partial (\Omega / \mathcal{V})}{\partial \mu_Y} = 0, \quad 
	\frac{\partial (\Omega / \mathcal{V})}{\partial \mu_T} = 0.
	\label{neutrality}
\end{equation}
Solving the system of equations \eqref{LandB} and \eqref{neutrality},
 with respect to the chemical potentials, one finds the desired relation
\begin{equation}
	\mu_\alpha = \omega_{\alpha \beta}(T) n_{\Delta_\beta} + \omega_B(T) n_B,
	\label{mu_gen}
\end{equation}
with  susceptibilities
\begin{equation}
	\omega(T) = \frac{1}{T^2} \left(
\begin{array}{ccc}
 a & b & b \\
 b &a & b \\
b & b & a \\
\end{array}
\right),\quad
a = \frac{22 \left(15 x^2+44\right)}{9 \left(17 x^2+44\right)},\quad 
b = \frac{8 \left(3 x^2+22\right)}{9 \left(17 x^2+44\right)},
\label{omega_ab}
\end{equation}
and
\begin{equation}
	\omega_B(T) = \frac{1}{T^2} \frac{4 \left(27 x^2+77\right)}{9 \left(17 x^2+44\right)},
	\label{omega_B}
\end{equation}
where $x = \langle \Phi(T) \rangle/T$.
The functions \eqref{omega_ab} and \eqref{omega_B} will be used later for numerical analysis
within  \textbf{Approach 2}.

In the standard treatment (i.e  \textbf{Approach 1} in our terms), one  has
to account for the redistribution of charges in plasma as well.
Above  $T_{sph}$, one can eliminate the chemical potential to baryon number
from \eqref{Omega} by applying the sphaleron constraint 
${\mu_B=-\frac13 \sum_\alpha \mu_\alpha}$.
In this case, the  chemical potentials to the lepton numbers are given 
by
\begin{equation}
	\mu_\alpha = \omega^{above}_{\alpha \beta}(T) n_{\Delta_\beta}, \quad T > T_{sph},
	\label{mu_above}
\end{equation}
with
\begin{equation}
\omega^{above}(T)=
 \frac{1}{T^2}
 \left(
\begin{array}{ccc}
 a' & b' & b' \\
 b' & a' & b' \\
 b' & b' & a' \\
\end{array}
\right),\quad
a' =  \frac{2 \left(963 x^2+2827\right)}{3 \left(333 x^2+869\right)},\quad b' = \frac{8 \left(55-9 x^2\right)}{3 \left(333 x^2+869\right)}.
\label{omega_ab_above}
\end{equation}

Below the sphaleron freeze-out, $B$ and $L$ are conserved
numbers within the SM so \eqref{mu_gen} is exact. The $n_B$ on the
r.h.s of \eqref{mu_gen} is fixed $n_B=n_{B^{eq}}(T_{sph})$. 
So, the lepton chemical potentials can now be written as
\begin{equation}
	\mu_\alpha = \omega_{\alpha \beta}(T) n_{\Delta_\beta} + \omega_B(T) 
	n_{B^{eq}}(T_{sph}),\quad T \leq T_{sph},
	\label{mu_below}
\end{equation}
with the same susceptibilities \eqref{omega_ab} and \eqref{omega_B}.
Note that the chemical potentials $\mu_\alpha$ calculated according to eqs. \eqref{mu_above} and \eqref{mu_below} are
 smooth functions of $\Delta_\alpha$ and the temperature.

% section susceptibilities (end)

\section{Kinetic equations for low-scale leptogenesis}
\label{nuMSM}
In this section we will present the set of kinetic equations
describing the production of lepton asymmetry in the extension
of the SM with two HNLs. This extension can be also viewed as a part of the $\nu$MSM
(neutrino minimal Standard Model) \cite{Asaka:2005an,Asaka:2005pn}.
For the convenience of the reader, we provide  relevant details
regarding the kinetic equations from ref. \cite{Eijima:2017anv} here.

The Lagrangian of the model is that of the type-I see-saw
with two right-handed neutrinos.
In what follows, however, 
it is more convenient
to use the pseudo-Dirac basis. So $N_2$ and $N_3$ are unified in one Dirac spinor 
$\Psi = N_2 + N_3^c$ \cite{Shaposhnikov:2008pf}.
The Lagrangian reads
\begin{equation}
\begin{aligned}
 \mathcal{L} &=  \mathcal{L}_{SM} +   \overline{\Psi} i \partial_\mu \gamma^\mu \Psi
 - M \overline{\Psi}\Psi +  \mathcal{L}_{int}, \\
\mathcal{L}_{int} &=
- \frac{\Delta M}{2} (\overline{\Psi}\Psi^c + \overline{\Psi^c}\Psi) 
 - (h_{\alpha 2} \langle \Phi \rangle \overline{\nu_{L \alpha}} \Psi + h_{\alpha 3} \langle \Phi \rangle \overline{\nu_{L\alpha}} \Psi^c + h.c.),
 \label{Lagrangian}
\end{aligned}
\end{equation}
where $\mathcal{L}_{SM}$ is the SM part, $M = (M_{3} + M_{2})/2$
is the common mass of HNLs and $\Delta M = (M_{3} - M_{2})/2$ is their Majorana mass difference.
The parametrization of the matrix of the neutrino Yukawa coupling constants $h_{\alpha I}$
is defined in appendix~\ref{sec:parametrization_of_yukawa_couplings}.

The system of kinetic equations describing oscillations
of two HNLs and their interactions with the SM particles was obtained in ref. \cite{Eijima:2017anv}
basing on the idea of separation of scales. The slow variables are
number densities $n_{\Delta_\alpha}$, as well as 
distribution functions and correlations of HNLs (and analogous for  anti-HNLs). 
The latter can be combined into the matrix $\rho_N$ ($\rho_{\bar{N}}$~for~anti-HNLs),
so that, e.g. the number density of the first HNL is 
${n_{N_1} = \int d k^3/(2\pi)^3 (\rho_N)_{1 1}}$. Introducing also the Fermi-Dirac
distribution function for massless neutrino ${f_\nu = 1/(e^{k/T}+1)}$ we can present
the equations of ref.~\cite{Eijima:2017anv} in the following form
\begin{subequations}
\begin{align}
i \frac{d n_{\Delta_\alpha}}{dt}
&= -  2 i \frac{\mu_\alpha}{T} \int \frac{d^{3}k}{(2 \pi)^{3}} \Gamma_{\nu_\alpha} f_{\nu} (1-f_{\nu})  \, 
    + i \int \frac{d^{3}k}{(2 \pi)^{3}} \left( \, \text{\text{Tr}}[\tilde{\Gamma}_{\nu_\alpha} \, \rho_{\bar{N}})
    -  \, \text{\text{Tr}}[\tilde{\Gamma}_{\nu_\alpha}^\ast \, \rho_{N}] \right),\label{kin_eq_a}
\\
i \, \frac{d\rho_{N}}{dt} 
&= [H_N, \rho_N]
    - \frac{i}{2} \, \{ \Gamma_{N} , \rho_{N} \} 
    - \frac{i}{2} \, \sum_\alpha \tilde{\Gamma}_{N}^\alpha \, \left[ 2 \frac{\mu_\alpha}{T} f_{\nu} (1-f_{\nu}) \right],\label{kin_eq_b}
\\
i \, \frac{d\rho_{\bar{N}}}{dt} 
&= [H_N^\ast, \rho_{\bar{N}}]
    - \frac{i}{2} \, \{ \Gamma_{N}^\ast , \rho_{\bar{N}} \} 
    + \frac{i}{2} \, \sum_\alpha (\tilde{\Gamma}_{N}^\alpha)^\ast \, \left[ 2 \frac{\mu_\alpha}{T} f_{\nu} (1-f_{\nu}) \right],
\label{kin_eq_c}
\end{align}
\label{kin_eq}
\end{subequations}
The main advantage of the approach of ref. \cite{Eijima:2017anv}
is that all quantities entering eqs. \eqref{kin_eq}
are expressed explicitly through the parameters of the Lagrangian \eqref{Lagrangian} and
functions which are calculable within the SM. Namely,
\begin{equation}
 H_N = H_0 + H_I, \quad
 H_0 = -\frac{\Delta M M}{E_N} \sigma_{1}, \quad
 H_I = h_{+} \sum_{\alpha} Y_{+, \alpha}^{N} + h_{-} \sum_{\alpha} Y_{-, \alpha}^{N},
\label{H_N}
\end{equation}
where $\sigma_1$ is the Pauli matrix, and
 \begin{equation}
\begin{aligned}
\Gamma_N &= \Gamma_+ + \Gamma_-,\quad
\Gamma_+ = \gamma_+ \sum_{\alpha} Y_{+, \alpha}^{N},\quad
% \label{ga-}
\Gamma_- = \gamma_- \sum_{\alpha} Y_{-, \alpha}^{N}, \;
 \tilde{\Gamma}_N^\alpha = - \gamma_{+} Y_{+, \alpha}^{N} + \gamma_{-} Y_{-, \alpha}^{N},\\
 \Gamma_{\nu_\alpha} &= (\gamma_+ + \gamma_-) \sum_I h_{\alpha I} h_{\alpha I}^\ast, \quad
 \tilde{\Gamma}_{\nu_\alpha} = - \gamma_{+, \alpha}^{\nu} Y_{+,\alpha}^{\nu}+ \gamma_{-, \alpha}^{\nu} Y_{-,\alpha}^{\nu}.
 \label{rates}
\end{aligned} 
\end{equation}
Here
\begin{equation}
\begin{aligned}
 h_+ &= \frac{2 \langle \Phi \rangle^2 E_\nu (E_N + k) (E_N + E_\nu)}
            {k E_N (4(E_N+E_\nu)^2 +\gamma_\nu^2)}, \quad
 h_- = \frac{2 \langle \Phi \rangle^2 E_\nu (E_N - k) (E_N - E_\nu)}
            {k E_N (4(E_N-E_\nu)^2 +\gamma_\nu^2)}, \\
 \gamma_+ &= \frac{2 \langle \Phi \rangle^2 E_\nu (E_N + k) \gamma_\nu}
            {k E_N (4(E_N+E_\nu)^2 +\gamma_\nu^2)}, \hspace{3em}
 \gamma_- = \frac{2 \langle \Phi \rangle^2 E_\nu (E_N - k) \gamma_\nu}
            {k E_N (4(E_N-E_\nu)^2 +\gamma_\nu^2)}, 
\end{aligned}
\end{equation}
while
\begin{equation}
\begin{aligned}
Y_{+, \alpha}^{N} &=
\begin{pmatrix}
  h_{\alpha 3} h_{\alpha 3}^\ast & - h_{\alpha 3} h_{\alpha 2}^\ast \\
 - h_{\alpha 2} h_{\alpha 3}^\ast & h_{\alpha 2} h_{\alpha 2}^\ast
\end{pmatrix}, \quad
Y_{-, \alpha}^{N} =
\begin{pmatrix}
  h_{\alpha 2} h_{\alpha 2}^\ast & - h_{\alpha 3} h_{\alpha 2}^\ast \\
 - h_{\alpha 2} h_{\alpha 3}^\ast & h_{\alpha 3} h_{\alpha 3}^\ast
\end{pmatrix}, \\
Y_{+, \alpha}^{\nu} &=
\begin{pmatrix}
  h_{\alpha 3} h_{\alpha 3}^\ast & - h_{\alpha 2} h_{\alpha 3}^\ast \\
 - h_{\alpha 3} h_{\alpha 2}^\ast & h_{\alpha 2} h_{\alpha 2}^\ast
\end{pmatrix}, \quad
Y_{-, \alpha}^{\nu} =
\begin{pmatrix}
  h_{\alpha 2} h_{\alpha 2}^\ast & - h_{\alpha 2} h_{\alpha 3}^\ast \\
 - h_{\alpha 3} h_{\alpha 2}^\ast & h_{\alpha 3} h_{\alpha 3}^\ast
\end{pmatrix},
\end{aligned}
\end{equation}
where  $E_\nu=k-b_L$ and $\gamma_\nu$ is the neutrino dumping rate. The function $b_L$---the
 neutrino potential in the medium---has
 been computed in a number of works (in different limits) \cite{Notzold:1987ik,Morales:1999ia}.
We use $\gamma_\nu$ and $b_L$ from ref. \cite{Ghiglieri:2016xye}. 

The system \eqref{kin_eq} 
is a set of differential and integro-differential equations.
The distribution functions on the l.h.s. of eqs. \eqref{kin_eq_b} and \eqref{kin_eq_c} 
depend on momentum
while on the r.h.s. of eq. \eqref{kin_eq_a} they are integrated over the  momentum.
This makes the solution of the system~\eqref{kin_eq} 
a technically involved and computational-resource consuming problem.
To simplify it, one can integrate equations~\eqref{kin_eq_b}~and~\eqref{kin_eq_c}   
over the momentum, accounting for the leading terms of the density matrices
\cite{Asaka:2011wq}.
These  are proportional to the equilibrium 
Fermi-Dirac distributions $f_N=1/(e^{E_N/T}+1)$.

It is also convenient to rearrange  equations \eqref{kin_eq}
by introducing the CP-even and CP-odd combinations
$\rho_+ \equiv (\rho_N+\rho_{\bar{N}})/2 - \rho_N^{eq}$,\quad  
$\rho_- \equiv \rho_N - \rho_{\bar{N}}$.
Integrated over momenta, they give the matrices
 $n_{+}$ and 
$n_{-}$, $n_{X} = \int dk^{3}/(2 \pi)^{3} \rho_{X}(k)$.
Integrating the effective Hamiltonian \eqref{H_N} and rates \eqref{rates}  over the
 $k_c = k/T$ and denoting
\begin{equation}
\begin{aligned}
 \overline{\Gamma}_{\nu_{\alpha}} &= \frac{6}{\pi^2} \int dk_c k_c^2 e^{k_c} f_\nu^2 \Gamma_{\nu_\alpha}, \quad
 \overline{\tilde{\Gamma}}_{\nu_{\alpha}} = \frac{T^3}{2 \pi^2} \frac{1}{n_N^{eq}} \int dk_c k_c^2 f_N \tilde{\Gamma}_{\nu_\alpha}, \\
 \overline{H}_N &= \frac{T^3}{2 \pi^2} \frac{1}{n_N^{eq}} \int dk_c k_c^2 f_N H_N, \quad
 \overline{\Gamma}_N = \frac{T^3}{2 \pi^2} \frac{1}{n_N^{eq}} \int dk_c k_c^2 f_N \Gamma_N, \\
 \overline{\tilde{\Gamma}}^{\alpha}_N &= \frac{6}{\pi^2} \int dk_c k_c^2 e^{k_c} f_\nu^2 \tilde{\Gamma}^{\alpha}_N, \quad
 S^{eq} = \frac{T^3}{2 \pi^2} \frac{1}{s} \int dk_c k_c^2 \dot{f_N}\cdot
 {\bf 1}_{2\times 2},
\end{aligned}
\end{equation}
we can present the kinetic equations \eqref{kin_eq} in the form
suitable for numerical computations
\begin{equation}
\begin{aligned}
 \dot{n}_{\Delta_{\alpha}} = &- \text{Re}\,\overline{\Gamma}_{\nu_{\alpha}} 
 \mu_\alpha\frac{T^2}{6} 
 + 2 i \text{Tr} [(\text{Im}\,\overline{\tilde{\Gamma}}_{\nu_{\alpha}})n_{+}] 
 - \text{Tr} [(\text{Re}\,\overline{\tilde{\Gamma}}_{\nu_{\alpha}}) n_{-}], \\
 \dot{n}_{+} = &-i [\text{Re}\,\overline{H}_N,n_{+}] + \frac{1}{2}[\text{Im}\,\overline{H}_N,n_{-}] 
 - \frac{1}{2}\{\text{Re}\,\overline{\Gamma}_N,n_{+}\} - \frac{i}{4} \{\text{Im}\,\overline{\Gamma}_N,n_{-}\} \\
 &  - \frac{i}{2} \sum (\text{Im}\,\overline{\tilde{\Gamma}}_N^{\alpha})\mu_\alpha\frac{T^2}{6} - S^{eq}, \\ 
 \dot{n}_{-} =&\; 2 [\text{Im}\,\overline{H}_N,n_{+}] -i [\text{Re}\,\overline{H}_N,n_{-}]
  - i\{\text{Im}\,\overline{\Gamma}_N,n_{+}\} - \frac{1}{2} \{\text{Re}\,\overline{\Gamma}_N,n_{-}\} \\
  &- \sum (\text{Re}\,\overline{\tilde{\Gamma}}_N^{\alpha}) \mu_\alpha\frac{T^2}{6}.
 \label{main_eq}
\end{aligned}
\end{equation}
Note that the system of kinetic equations in the symmetric phase has exactly the same
 form as \eqref{main_eq}, but the expressions for the effective Hamiltonian~\eqref{H_N} and
 the rates~\eqref{rates} have to be modified \cite{Eijima:2017anv}.

The above system  \eqref{main_eq}, together with  \eqref{kine_ec_B},
form the basis for the numerical analysis which we will perform in the next section.

\section{Analysis of kinetic equations} % (fold)
\label{sec:analysis_of_kinetic_equations}

In this section we will numerically solve the kinetic equations in both approaches
described in section \ref{sec:structure_of_kinetic_equations}.
This will enable us to examine the accuracy of the usual assumption concerning 
the instantaneous freeze-out of the sphalerons.

Let us summarize the differences between the two approaches:
\begin{enumerate}
\item \textbf{Approach 1}. Instantaneous $B$ freeze-out. 

The baryon number density  is
${n_B(T) = n_{B^{eq}}(T)}$, for all 
temperatures above $T_{sph}$, while 
\newline
${n_B(T) = n_{B^{eq}}(T_{sph})}$ for all $T\leq T_{sph}$.

One has to solve eqs. \eqref{main_eq}
with the chemical potentials $\mu_\alpha$ defined by eqs.
\eqref{mu_above} and \eqref{omega_ab_above} at $T>T_{sph}$ 
and by eqs.  \eqref{omega_ab}, \eqref{omega_B} and \eqref{mu_below} at $T \leq T_{sph}$.

\item \textbf{Approach 2}. Inclusion of a separate kinetic equation 
for  $n_B$.

In this case, one can follow the  $n_B$ during the freeze-out,
 but at the cost of adding a new time-scale into the problem.

One has to solve the system of eqs. \eqref{main_eq} and eq. \eqref{kine_ec_B}
with the chemical potentials $\mu_\alpha$ defined by eqs.
\eqref{mu_gen}, \eqref{omega_ab} and \eqref{omega_B}.
\end{enumerate}

So far we have not accounted for the expansion of the Universe.
This can be easily achieved by changing variables so that all densities
densities in eqs. \eqref{main_eq}
and \eqref{B_eq} are  divided
by the entropy density $s(T)$. For the numerical computations
we used $s(T)$ calculated in refs. \cite{Laine:2006cp,Laine:2015kra}
in the various temperature regions.
We also change the argument from time to temperature (to be more precise,
we use $z = \log (M/T)$, but all graphs are presented as functions of $T$).
\newline
It is convenient to denote $\Delta(T) = \sum_\alpha n_{\Delta_\alpha(T)}/s(T)$,
since this is the quantity which is related to the baryon asymmetry by sphaleron processes.
In what follows, by the baryon asymmetry we mean  $B(T)= n_B(T)/s(T)$.
Note that we are  interested  in the behaviour of $B(T)$ in  general, rather than
particular values of $B$.
A full scan of the parameter space is required to determine the values of the parameters
responsible for the correct value of $B$. We leave this for future work.

We solve both systems with  zero initial conditions.\footnote{See ref. \cite{Asaka:2017rdj} for
a discussion the impact of initial conditions on BAU.}
In order to determine an appropriate initial temperature, it is useful to note
that HNL oscillations begin (see, e.g. \cite{Akhmedov:1998qx,Asaka:2005pn}) roughly at temperature  
\begin{equation}
	T_{osc} \simeq \l \frac{\Delta M M M_0}{3} \r^{1/3}.
\end{equation}
We start from an initial temperature at which
asymmetries are  negligible (i.e. the initial temperature is much higher than $T_{osc}$)
 and solve the equations down to $T=100~$GeV.\footnote{
We solve both systems in Wolfram Mathematica, using the NDSolve function
which is able to handle stiff systems properly.
}
For the purposes of the present work we have fixed most of the parameters and varied only 
the mass difference $\Delta M$ as well as the imaginary part of the complex mixing angle $X_\omega$,  
see appendix~\ref{sec:parametrization_of_yukawa_couplings} for details.
 These values are listed in table \ref{tab:param}.

In both approaches, we explicitly compute the asymmetry $\Delta(T)$. 
Therefore, it is natural to check 
whether these functions coincide in the two approaches.
Our numerical results show that this is  indeed the case.
For temperatures above $100$~GeV, the relative deviation of these quantities 
in the two approaches does not exceed $10^{-4}$ for all the parameter sets that we have checked. 
Moreover, this deviation is 
saturated mostly by numerical errors.

\begin{figure}[htb!]
\centering % \begin{center}/\end{center} takes some additional vertical space
\includegraphics[width=.85\textwidth]{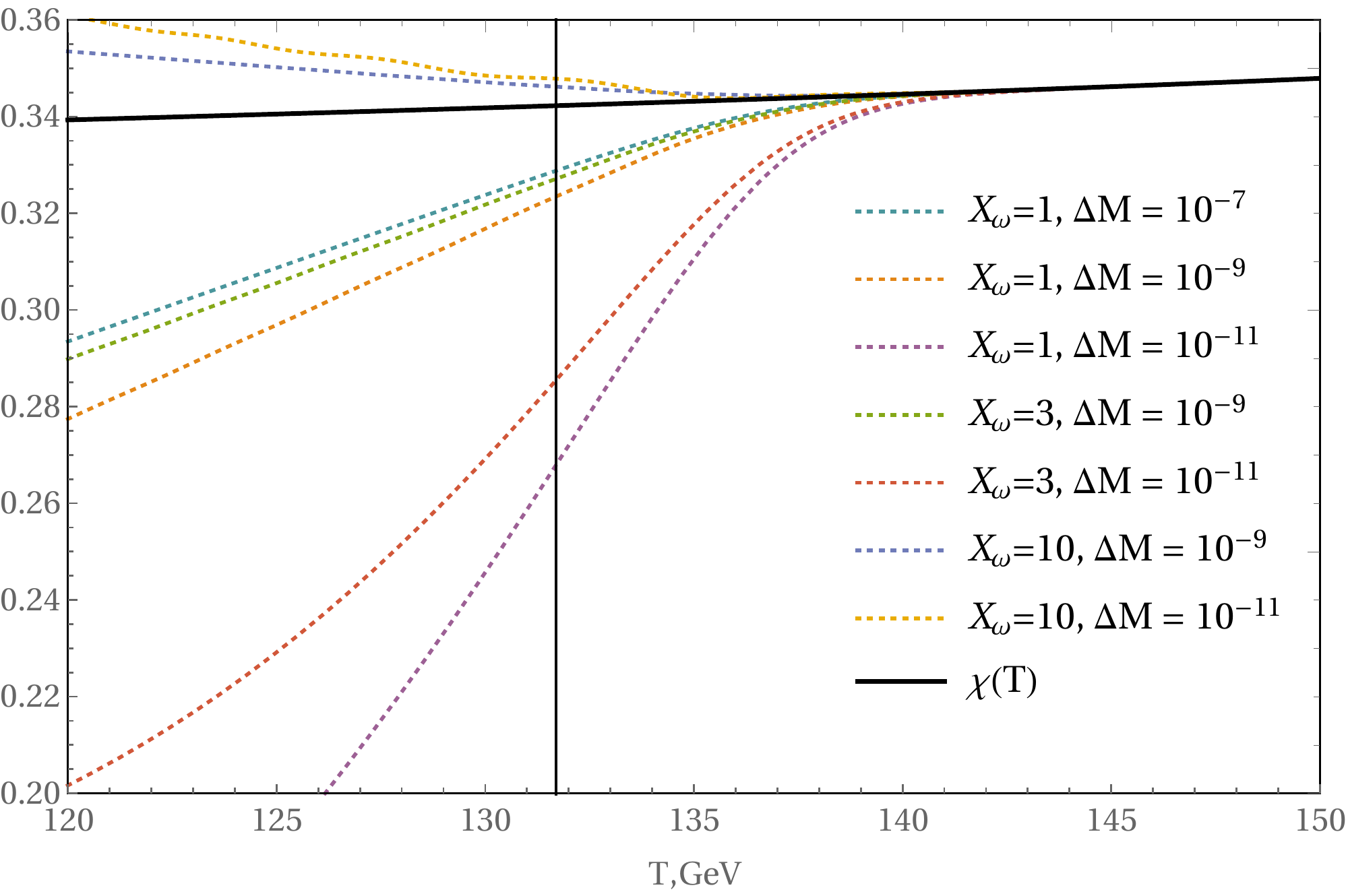}\\

\includegraphics[width=.85\textwidth]{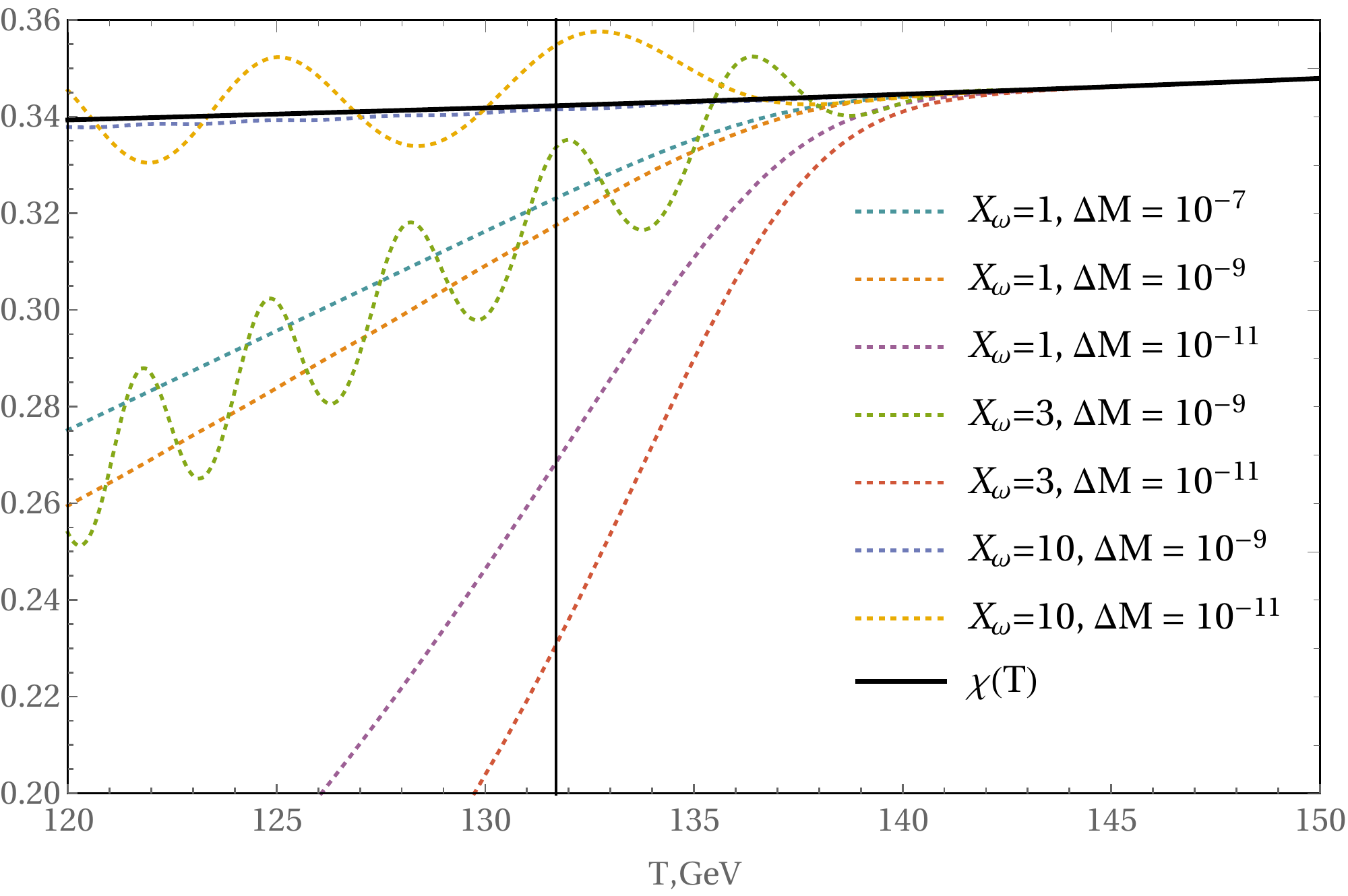}
% "\includegraphics" is very powerful; the graphicx package is already loaded
\caption{\label{equil} The ratio $r(T) = -B(T)/\Delta(T)$ as function of temperature
for the inverted (upper panel) and normal (lower panel) hierarchies.
The solid black line represents $\chi(T)$. In the equilibrium with respect to sphalerons,
$r(T) = \chi(T)$.
The thin black vertical line shows the sphaleron freeze-out temperature 
$T_{sph}=131.7$~GeV. The common mass of HNLs is $M=1$~GeV. The other parametes are specified in 
table~\ref{tab:param}.}
\end{figure}

Now that we have shown that both approaches are capable of evaluating the value of $\Delta(T)$ consistently,
 we can study how accurate \textbf{Approach 1}  can be 
in what concerns the value of $B$.
While calculating \eqref{mu_above}, we have assumed the validity of \eqref{B_eq} at all temperatures
down to $T_{sph}$. One can check  this assumption using $B(T)$ and $\Delta(T)$ computed in 
\textbf{Approach 2}.
To this end, one can define
$r(T)=-B(T)/\Delta(T)$. If the system is in equilibrium with respect
to the sphaleron processes, the value of this ratio should coincide with $\chi(T)$ defined in \eqref{B_eq}.
We show the behaviour of $r(T)$ for different values of $\Delta M$ and $X_\omega$ in 
figure~\ref{equil}.
It is clear that $r(T)$
ceases to follow $\chi(T)$ at temperatures above the freeze-out temperature
$T_{sph}$.
The different shape of the curves in figure \ref{equil} is due to the different regimes
of the asymmetry generation in $\Delta_\alpha$. 
The important universal feature of all the presented curves is
that the temperature at which they deviate from $\chi(T)$ is $T_{dev}\simeq 140~$GeV.

\begin{figure}[htb!]
\centering % \begin{center}/\end{center} takes some additional vertical space
\includegraphics[width=.85\textwidth]{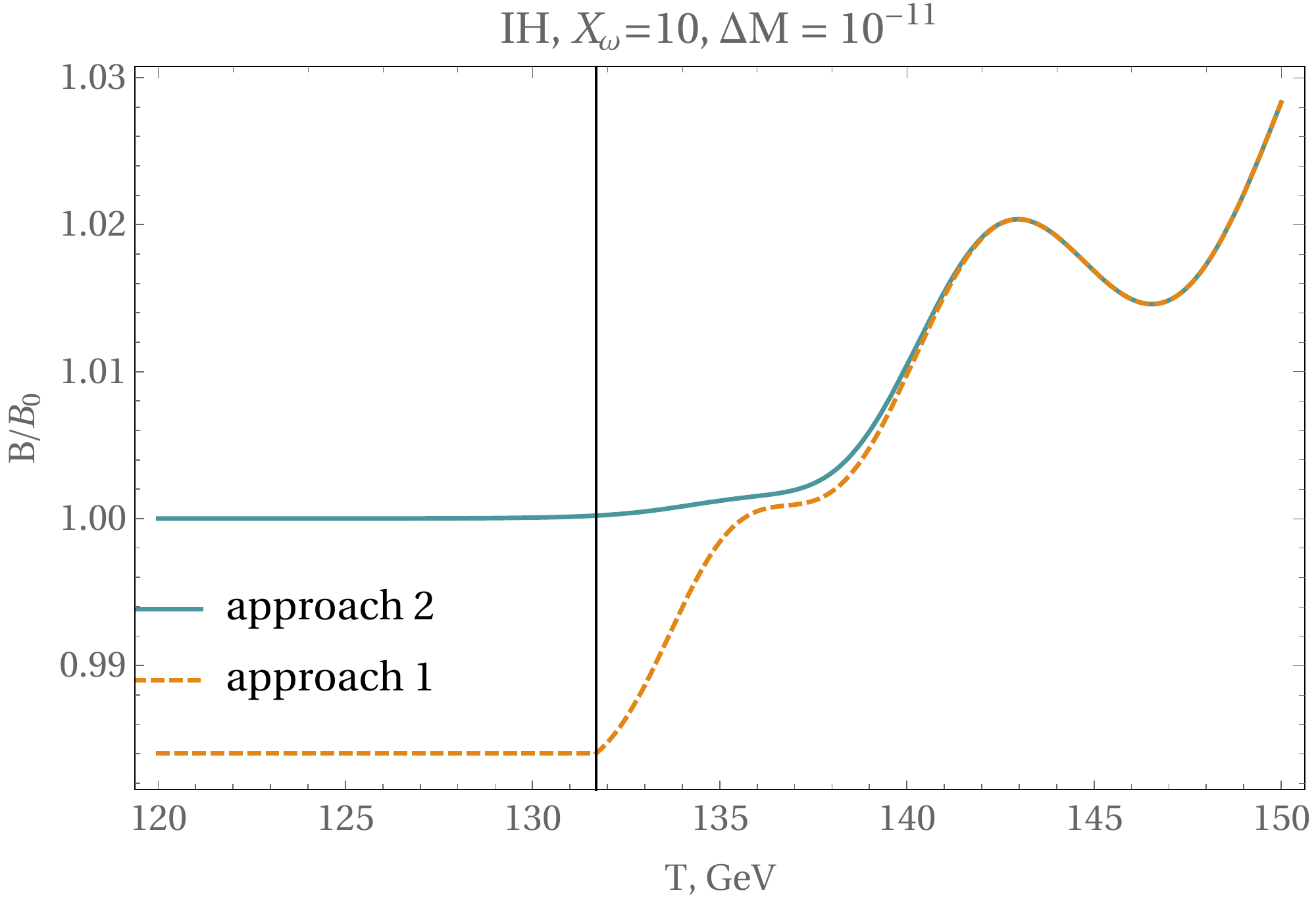} \\

\includegraphics[width=.85\textwidth]{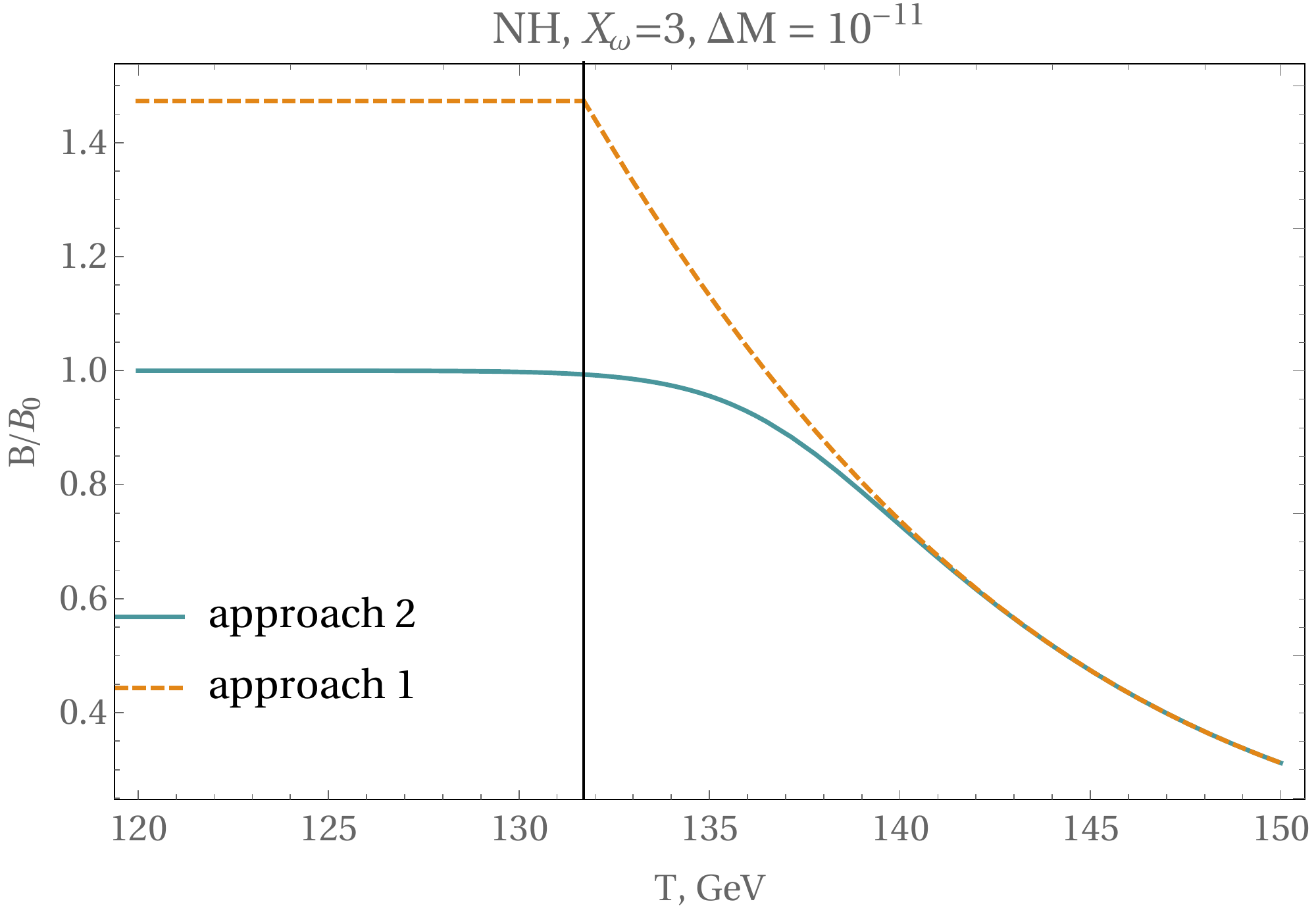}
% "\includegraphics" is very powerful; the graphicx package is already loaded
\caption{\label{Bfig} $B(T)/B_0$ as a function of temperature
in  the approach with instant freeze-out (dotted orange line) and in the approach 
with the separate equation for $B$ (blue line).
Upper panel: inverted hierarchy, lower panel: normal hierarchy.
The thin black vertical line shows the sphaleron freeze-out temperature 
$T_{sph}=131.7$~GeV. The common mass of HNLs is $M=1$~GeV.
The rest of the parameters are specified in 
table~\ref{tab:param}.
See also the discussion in the main text. 
}
\end{figure}

To understand better the behaviour of the system at temperatures around $130$~GeV,
we can also trace the evolution of the baryon asymmetry $n_B$ itself.
This is shown in figure~\ref{Bfig}.
Let us denote  $B_0 = B^\text{appr.2}(100\text{~GeV})$ in  
\textbf{Approach 2}. We will assume that this is the eventual
value of BAU.
The dashed orange line in figure~\ref{Bfig} shows $B^\text{appr.1}(T)/B_0$,
while the solid blue line shows $B^\text{appr.2}(T)/B_0$.

We have tested different combinations of $\Delta M$ and $X_\omega$
in the intervals from table \ref{tab:param}.
For the most of the parameter sets, the deviation of ${B^\text{appr.1}(T)/B_0}$ from $1$
is typically small, see table~\ref{tab:IH}.
All cases share the same dynamics of the freeze-out.
First, at temperatures around $T_{dev}\simeq140$~GeV there is a deviation 
from equilibrium. The final freeze-out occurs at temperatures 
around $130$~GeV. 

\begin{table}[htb!]
\centering
\begin{tabular}{|l|c|c|c|c|}
\hline
$\Delta M$ \textbackslash $X_\omega$ & 1 & 3 & 5 & 10\\
\hline
$10^{-7}$ & $1.04$ & $1.03$ & $1.01$ & $0.98$\\
\hline
$10^{-8}$ & $1.05$ & $1.03$ & $1.01$ & $0.98$\\
\hline
$10^{-9}$ & $1.06$ & $1.05$ & $1.02$ & $0.99$\\
\hline
$10^{-10}$ & $1.07$ & $1.06$ & $1.03$ & $0.99$\\
\hline
$10^{-11}$ & $1.27$ & $1.20$ & $1.02$ & $0.98$\\
\hline
\end{tabular} \quad
\begin{tabular}{|l|c|c|c|c|}
\hline
$\Delta M$ \textbackslash $X_\omega$ & 1 & 3 & 5 & 10\\
\hline
$10^{-7}$ & $1.06$ & $1.04$ & $1.02$ & $0.99$\\
\hline
$10^{-8}$ & $1.06$ & $1.04$ & $1.03$ & $0.99$\\
\hline
$10^{-9}$ & $1.08$ & $1.02$ & $1.03$ & $1.00$\\
\hline
$10^{-10}$ & $1.11$ & $1.17$ & $1.11$ & $0.99$\\
\hline
$10^{-11}$ & $1.27$ & $1.47$ & $0.96$ & $0.97$\\
\hline
\end{tabular}
\caption{\label{tab:IH} The ratio ${B^\text{appr.1}(100\text{~GeV})/B_0}$ for 
different values of $\Delta M$ and $X_\omega$. All other parameters 
are fixed, see appendix \ref{sec:parametrization_of_yukawa_couplings}.
Left table: inverted hierarchy, right table: normal hierarchy.
For larger values of $X_\omega$ the deviations are also quite small.}
\end{table}
As one can see from  table \ref{tab:IH}, the 
approximation of the instantaneous freeze-out works quite good. The reason is that
for the values of the parameters from table \ref{tab:param} the amount
of the asymmetry $\Delta$ generated during the transition period is  not very large.

\begin{figure}[htb!]
\centering
\includegraphics[width=.865\textwidth]{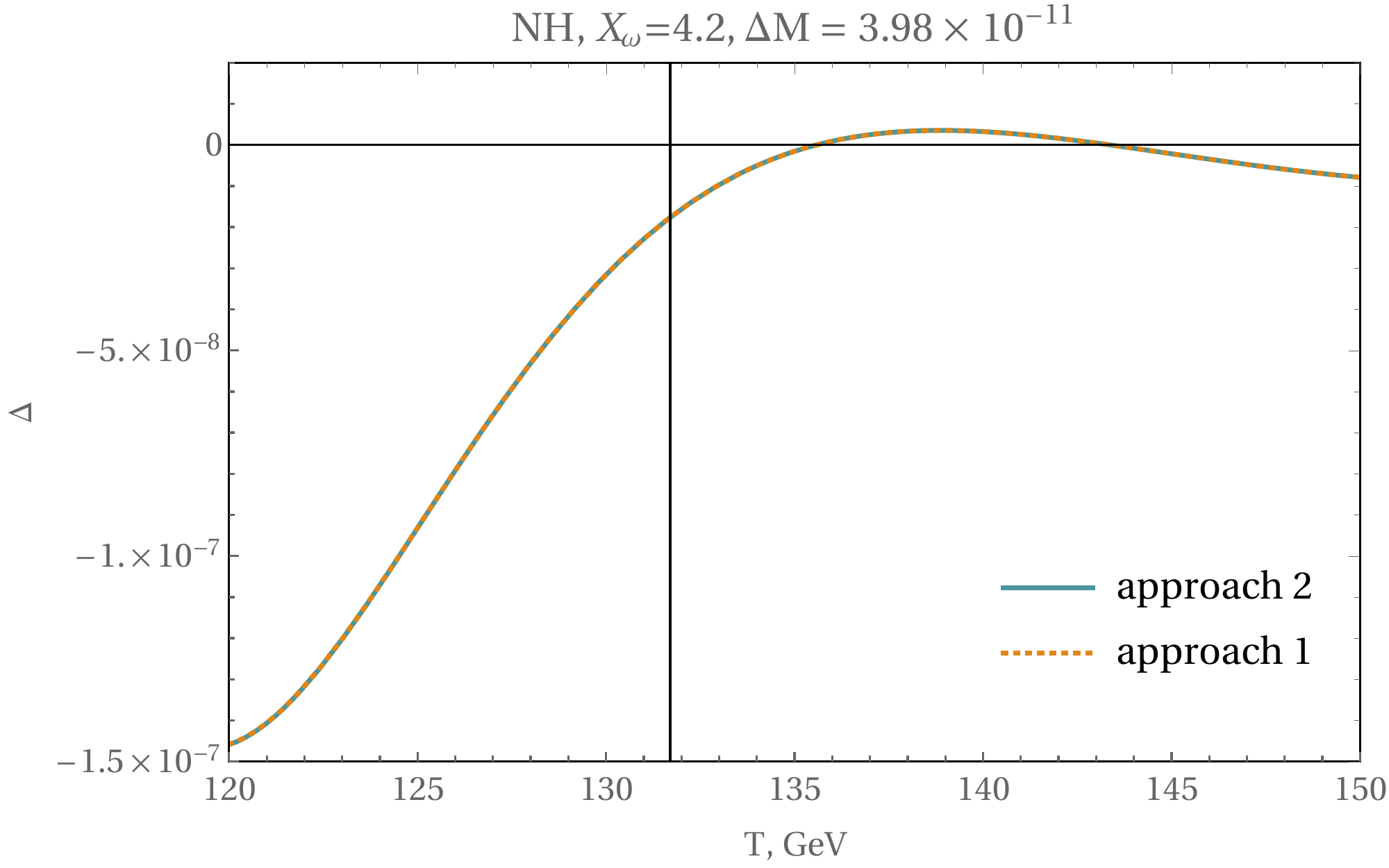}\\

\hspace{0.5em}\includegraphics[width=.85\textwidth]{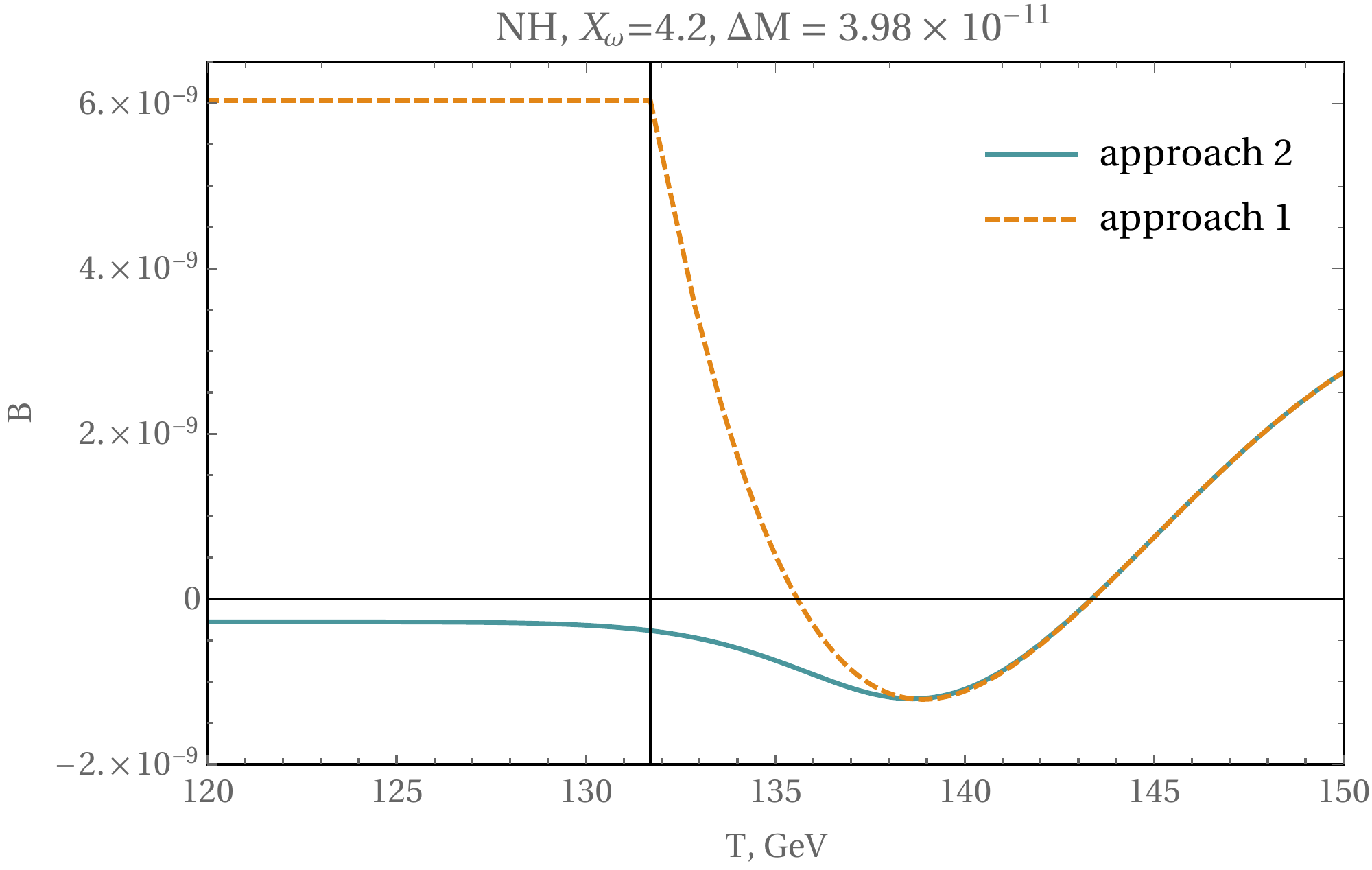} 
\caption{\label{huge_deviation} 
An example of a large deviation. $B^\text{appr.1}/B^\text{appr.2} \simeq -21.9$.
Upper panel, $\Delta(T)$ as function of temperature
in the approach with instant freeze-out (dotted orange line) and in the approach 
with the separate equation for $B$ (blue line). 
Lower panel, $B(T)$ as function of temperature.
The black vertical line shows the sphaleron freeze out temperature 
$T_{sph}=131.7$~GeV. 
To illustrate a situation when lepton asymmetry is generated right before the sphaleron freeze-out
we utilize the following values of the parameters:
the common mass of HNLs is $M=1$~GeV,
$\Delta M = 3.98\cdot10^{-11}$~GeV, $X_\omega = 4.2$,
$\delta = 3 \pi /2 $, $\eta = 19 \pi /16$, $\mathrm{Re}\,\omega =\pi/4$.
See also the discussion in the main text. 
}
\end{figure}

Still, the difference between $B(T)$ in the two approaches can be quite large for some parameters.
We illustrate this in figure \ref{huge_deviation}.
As one can see from the upper panel of the figure,
this is precisely the case when a sufficient portion of the 
 asymmetry $\Delta(T)$ is generated
in the temperature interval between $130$~GeV and $140$~GeV,
when the sphalerons are not fast enough to redistribute 
all the generated asymmetry.
Moreover, $\Delta(T)$ even changes its sign in this interval and while the actual 
$B(T)$ remains negative,  the $B^\text{appr.1}(T_{sph})$ is positive.

\section{Improvement of the instant freeze-out approach} % (fold)
\label{sec:improvement_of_the_instant_freeze_out_approach}
In the previous section, we have shown that both approaches coincide 
for the  lepton asymmetry, but the value of BAU
differs.
In principle, one can implement \textbf{Approach 2} in order to
ensure that the eventual value of BAU is correct for 
any set of the model parameters.
This requires, however, an introduction of the new time scale $\Gamma_B(T)^{-1}$
to a stiff problem.
In this section, we will present a resolution 
to this complication.

In contrast to $\Delta(T)$, $B(T)$ does not change sufficiently
after the deviation from the equilibrium at ${T_{dev}\simeq 140}$~GeV
since the sphaleron rate decreases.
Therefore, at first sight it is reasonable to take $B=B^{eq}(T^*)$ at some
temperature $T^*$ between $T_{dev}$ and $T_{sph}$.
This simple recipe works to some extent, but it is not universal, since the optimal
value of $T^*$ depends on the regime of the asymmetry generation.

We have found that there exists a universal solution which does not
require any modifications to the main system of kinetic equations.
One can actually benefit from the fact that both approaches coincide
with regard to the produced amount of $\Delta$.
Let us rewrite the kinetic equation for $B$ as
\begin{equation}
	\frac{d (B(T))}{d T} \frac{d T}{d t} = -\Gamma_B(T)
	 \left (B(T) + \chi(T) \Delta(T) \right),
\label{kin_B_again}
\end{equation}
where  $\chi(T)$ was defined in eq. \eqref{B_eq},
and  the temperature dependence is shown explicitly.
Previously, in \textbf{Approach~2}, we  included this equation into the full
system of kinetic equations. However, one can adopt another point of view 
for eq. \eqref{kin_B_again}: it is a single first order differential equation
with a temperature dependent source 
$\chi(T)\Delta(T)$. The source as a function of $T$
can be obtained within  \textbf{Approach 1} and then plugged into 
eq.~\eqref{kin_B_again}.
It is natural to expect that the solution of this equation will coincide with
the value of $B(T)$ obtained within \textbf{Approach 2}.
We have solved eq.~\eqref{kin_B_again} numerically and found that this is indeed the case
for all regimes of the asymmetry generation, including the case of figure \ref{huge_deviation}.
The deviation between $B(T)$ in the improved \textbf{Approach 1} and \textbf{Approach 2}
is of the same order as that of $\Delta(T)$.
It is enough to solve eq.~\eqref{kin_B_again} in the Higgs phase,
starting from the temperature that is higher than $T_{dec}\simeq 140$~GeV.
 We take $T_{in} = 150$~GeV.
At this temperature, the sphalerons are fast enough so there is actually no dependence
on the initial condition  $B(T_{in})$, namely, any initial value of $B$
will be immediately driven to the equilibrium value $\chi(T) \Delta(T)$.

To summarize, instead of using formula \eqref{standard_formula}
for the resulting BAU in numerical calculations,
we can reach very precise values by
solving eq. \eqref{kin_B_again} with the source term obtained
within  \textbf{Approach 1}.

% section improvement_of_the_instant_freeze_out_approach (end)

\section{Conclusions} % (fold)
\label{sec:conclusions}

In this paper we  studied the production of baryon asymmetry in 
low-scale leptogenesis.
Our main goal was to examine the accuracy of the standard approach
of  the BAU calculation based on the equilibrium relation \eqref{standard_formula}.
First, we clarified the proper usage of this equation \eqref{standard_formula}.
Namely, we described how one can  account for the neutrality of the electroweak plasma,
for the temperature dependence of the Higgs vacuum expectation value and, eventually, for
the prompt freeze-out of the sphaleron processes.
All these phenomena are  encoded in the relations
between chemical potentials to lepton numbers $\mu_\alpha$ and $n_{\Delta_\alpha}$, see
eqs. \eqref{mu_above}, \eqref{omega_ab_above}, \eqref{mu_below} 
and \eqref{omega_ab}.
The results obtained this way were compared to the kinetic description
of the $B$ freeze-out.
\newline
Solving numerically the kinetic equations for both approaches we have found:
\begin{itemize}
	\item The asymmetries $\Delta(T)$ as functions of $T$ perfectly coincide
	in both approaches.
	\item In contrast to the instant freeze-out assumption, the baryon number
    freeze-out occurs in two steps. At temperature
	$T_{dev}\simeq 140~$GeV, the baryon asymmetry starts to deviate from the equilibrium
	value described by eq. \eqref{standard_formula}. At temperatures around
	$T_{sph}\simeq 131.7$~GeV, the baryon asymmetry finally freezes out.
	\item If the production rate of lepton asymmetry was large during the transition
	period before $T_{dev}$ and $T_{sph}$, the final value of the baryon asymmetry
	deviates from the equilibrium one.
\end{itemize}
Therefore, for a comprehensive study of the parameter space of the model,
one has to account for the sphaleron transitions at temperatures close to 
$T_{sph}$. This can be done within  \textbf{Approach 1}, by solving 
the kinetic equation for the baryon asymmetry \eqref{kin_B_again}
with a source term $\sum_\alpha n_{\Delta_\alpha(T)}/s(T)$.
Such an improved procedure does not require modification of the main system of kinetic equations
and reproduces the correct results.

We studied the system of kinetic equations for two HNLs. However, our results apply to any scenario 
in which generation of the lepton asymmetry takes places at temperatures comparable with  $T_{sph}$.

% section conclusions (end)

\acknowledgments
We thank Jacopo Ghiglieri, Georgios Karananas, Mikko Laine, and Sander Mooij
 for helpful comments on the paper.
This work was supported by the ERC-AdG-2015 grant 694896. 
The work of  M.S., and I.T. was supported partially by the Swiss National Science Foundation.

\appendix
\section{Parametrization of Yukawa couplings} % (fold)
\label{sec:parametrization_of_yukawa_couplings}

In the type-I see-saw mechanism 
one can parametrize the neutrino Yukawa coupling constants for the mass basis of HNLs, denoted by $F_{\alpha I}$, 
in a way that allows to fix
the active neutrino mixing angles and masses from the experimental data.
It is the so-called Casas-Ibarra parametrization \cite{Casas:2001sr}. 
For more details see ref.~\cite{Asaka:2011pb}.

In the minimal case with two right-handed neutrinos 
the matrix of the Yukawa coupling constants $F_{\alpha I}$ can be parametrized as follows
\begin{equation}
	F = \frac{i}{v_{0}} U m_\nu^{1/2} \Omega m_N^{1/2},
\end{equation}
where $v_{0} = 174.1$ GeV, $m_\nu$ and $m_N$ are diagonal mass matrices for the three active neutrinos and the two HNLs. The
PMNS matrix $U$ is given by
\begin{equation}
U =  \left(
\begin{array}{ccc}
 c_{12} c_{13} & c_{13} e^{i \eta } s_{12} & e^{-i \delta } s_{13} \\
 -c_{23} s_{12}-c_{12} e^{i \delta } s_{13} s_{23} & e^{i \eta } \left(c_{12} c_{23}-e^{i \delta }
   s_{12} s_{13} s_{23}\right) & c_{13} s_{23} \\
 s_{12} s_{23}-c_{12} c_{23} e^{i \delta } s_{13} & -e^{i \eta } \left(c_{12} s_{23}+c_{23} e^{i
   \delta } s_{12} s_{13}\right) & c_{13} c_{23} \\
\end{array}
\right),
\end{equation}
(we use the central values for the mixing angles and squared mass differences 
from the global analysis of oscillation data \cite{Esteban:2016qun}),
and the orthogonal matrix $\Omega$ reads
\begin{align}
\Omega &=	\left(
\begin{array}{cc}
 0 & 0 \\
 \cos \omega &  \sin \omega \\
 -\xi  \sin \omega & \xi  \cos \omega \\
\end{array}
\right)\quad\text{for NH},\\
\Omega &=	\left(
\begin{array}{cc}
 \cos \omega &  \sin \omega \\
 -\xi  \sin \omega & \xi  \cos \omega \\
 0 & 0 \\
\end{array}
\right)\quad\text{for IH},
\end{align}
with a complex mixing angle $\omega$. The imaginary part of $\omega$ can be expressed as
\begin{equation}
	X_\omega=\exp( \mathrm{Im}\, \omega).
\end{equation}

The value of $X_\omega$ plays an important role 
in the dynamics of the asymmetry production since it determines the scaling of the
Yukawa couplings 
(see e.g. the discussion in section 3 of ref. \cite{Eijima:2017anv}) and, 
consequently, it controls the rates in eq. \eqref{rates}.

The neutrino Yukawa coupling constants in the pseudo-Dirac basis, $h_{\alpha I}$, 
entering eq. \eqref{Lagrangian} can be obtained by  the following transformation
\begin{equation}
 h = F U_{N}^{t},
\end{equation}
with a unitary matrix
\begin{equation}
 U_{N} = \frac{1}{\sqrt{2}} \left(
\begin{array}{cc}
 - i &  1 \\
 i & 1\\
\end{array}
\right).
\end{equation}

The values of the parameters which we use to produce most of the plots in this study are listed in table~\ref{tab:param}.
\begin{table}[h]
\centering
\begin{tabular}{|c|c|c|c|c|c|c|}
\hline
 $M$, GeV & $\Delta M$, GeV&$\delta$ & $\eta$ & $\mathrm{Re}\, \omega$ &$X_\omega$&$\xi$\\
% $\delta$ & $\eta$ & $\mathrm{Re}\, \omega$ & $\xi$  & $\Delta M$, GeV& $X_\omega$& $M$, GeV\\
\hline
$1$ & $10^{-7} - 10^{-11}$ & $\frac32 \pi$  & $\frac14 \pi$ & $\frac18 \pi$ &$1 - 10$& 1\\
\hline
\end{tabular}
\caption{\label{tab:param} Values of the parameters which have been used
for generating the figures \ref{equil} and \ref{Bfig}, as well as table \ref{tab:IH}.}
\end{table}
% section parametrization_of_yukawa_couplings (end)

% % This is the most common positions for acknowledgments. A macro is
% % available to maintain the same layout and spelling of the heading.

% % \paragraph{Note added.} This is also a good position for notes added
% % after the paper has been written.

% % The bibliography will probably be heavily edited during typesetting.
% % We'll parse it and, using the arxiv number or the journal data, will
% % query inspire, trying to verify the data (this will probalby spot
% % eventual typos) and retrive the document DOI and eventual errata.
% % We however suggest to always provide author, title and journal data:
% % in short all the informations that clearly identify a document.

\bibliographystyle{JHEP}
\bibliography{SphaleronsRefs}
\end{document}